\documentstyle[12pt,epsf]{article}
\catcode`\@=11

\begin{document}
\begin{titlepage}
\samepage{
\setcounter{page}{1}
\rightline{JHU-TIPAC-97011}
\rightline{July, 1997}
\vfill
\vfill
\begin{center}
{\Large \bf QCD Corrections to Flavor-Changing Neutral Currents
in the Supersymmetric Standard Model\\}
\vfill
\vfill
 {\large Jonathan A. Bagger, Konstantin T. Matchev
\footnote{Address after Sept. 1, 1997: Theory Group, Fermi National
Accelerator Laboratory, P.O. Box 500, Batavia, IL 60510.}
and Ren-Jie Zhang
\footnote{Address after Sept. 1, 1997: Department of Physics, University
of California, Davis, CA 95618 and Department of Physics, University of
Wisconsin, Madison, WI 53706.}\\}
\vspace{.25in}
 {\it  Department of Physics and Astronomy\\
       The Johns Hopkins University\\
       Baltimore, Maryland 21218 USA\\}
\end{center}
\vfill
\vfill
\begin{abstract}
{\rm
We compute the leading QCD corrections to $K$-${\overline K}$
mixing in the supersymmetric standard model with
general soft supersymmetry-breaking parameters.  We construct
the $\Delta S=2$ effective Lagrangian for three hierarchies of
supersymmetric particle masses, namely, when the gluino mass
is comparable to, much greater than, or much less than the
masses of the first two generation squarks.  We find that
the QCD corrections tighten the limits on squark mass
splittings by more than a factor of two.}
\end{abstract}
\vfill}
\end{titlepage}


\catcode`@=11
\long\def\@caption#1[#2]#3{\par\addcontentsline{\csname
  ext@#1\endcsname}{#1}{\protect\numberline{\csname
  the#1\endcsname}{\ignorespaces #2}}\begingroup
    \small
    \@parboxrestore
    \@makecaption{\csname fnum@#1\endcsname}{\ignorespaces #3}\par
  \endgroup}
\catcode`@=12

\newcommand{\newc}{\newcommand}
\newc{\gsim}{\lower.7ex\hbox{$\;\stackrel{\textstyle>}{\sim}\;$}}
\newc{\lsim}{\lower.7ex\hbox{$\;\stackrel{\textstyle<}{\sim}\;$}}
\def\tr{\mathop{\rm tr}}
\def\Tr{\mathop{\rm Tr}}
\def\Im{\mathop{\rm Im}}
\def\Re{\mathop{\rm Re}}
\def\bR{\mathop{\bf R}}
\def\bC{\mathop{\bf C}}
\def\lie{\mathop{\hbox{\it\$}}} 
\newc{\Qeff}{Q_{\rm eff}}
\newc{\mz}{M_Z}
\newc{\vphi}{\varphi}
\newc{\ve}{\varepsilon}
\newc{\ol}{\overline}
\newc{\GeV}{{\rm GeV}}
\newc{\TeV}{{\rm TeV}}
\newc{\re}{{\rm Re}}

\newc{\npb}{{\it Nucl. Phys. }B\ }
\newc{\plb}{{\it Phys. Lett. }B\ }
\newc{\prd}{{\it Phys. Rev.  }D\ }
\newc{\prl}{{\it Phys. Rev. Lett.\ }}
\newc{\pr}{{\it Phys. Rep.\ }}
\newc{\mpla}{{\it Mod. Phys. Lett. }A\ }
\newc{\zpc}{{\it Z. Phys. }C\ }
%



\section{Introduction} \label{sec:intro}

Low-energy supersymmetry is a leading candidate for physics beyond
the standard model because it stabilizes the gauge hierarchy.
However, the minimal supersymmetric extension to the standard model
contains over 100 new parameters.  Present experimental searches are
beginning to place significant constraints on the parameter space.
To date, the most important limits arise from the study of
flavor-changing neutral currents (FCNC).

In general, supersymmetric particles give rise to large contributions
to flavor-changing neutral currents \cite{longlist}.  The present limits
on these processes place strict constraints on the squark and slepton
mass matrices.  Indeed, the discovery of supersymmetric FCNC will be
an important step towards understanding the source of supersymmetry
breaking.

There are several FCNC processes which constrain the parameters of
supersymmetric models.  For example, the rare decay $b \to s\gamma$
has been used to constrain the mass parameters of the third generation
squarks \cite{list}.  In this paper, we will focus on $K$-${\overline
K}$ mixing, which gives the most stringent limit on squark masses of
the {\it first two} generations.

In what follows, we will extend the analysis of
Refs.~\cite{old}-\cite{ggms}
to include the leading order QCD corrections to $K$-${\overline
K}$ mixing in the minimal supersymmetric standard model (MSSM).
The QCD corrections are clearly very
important, and for the standard model, complete leading order
\cite{GW} and next-to-leading order \cite{NLO} analyses have
been carried out.  (Our results can be easily generalized to
$B$-${\overline B}$ and $D$-${\overline D}$ mixing,
in which case they also involve the third generation squarks.)

For general squark masses, the most important supersymmetric
contribution to $K$-${\overline K}$ mixing comes from the gluino box
diagrams.\footnote{In the minimal supergravity model, the chargino
and neutralino box diagrams can be of comparable importance
\cite{longlist}.  For a recent review, see Ref.~\cite{pokmis}.}
We will, therefore, restrict our attention to the gluino diagrams,
and ignore all diagrams with chargino and neutralino exchanges.
These diagrams depend on the details of the Higgs sector and are
generally less important than the diagrams we consider here.

As is by now a standard procedure, we define the source of FCNC in
terms of dimensionless flavor-changing insertions $\delta$, which
parametrize small deviations from the case of flavor-diagonal soft
squark masses.  We follow the notation of Ref.~\cite{ggms}, but we
omit the generation and weak isospin labels because we concentrate
solely on the first two generations.

This paper is organized as follows.  In Sec.~\ref{form} we review
the general formalism for calculating QCD corrections in the context
of effective field theory.  We then consider three cases for the
hierarchy of the superparticle masses: (a) $M_{\tilde g}
\simeq M_{\tilde q},$ where the gluino mass, $M_{\tilde g}$, is on
the order of the average squark mass, $M_{\tilde q}$, as well as
(b) $M_{\tilde g} \ll M_{\tilde q}$ and (c) $M_{\tilde g} \gg
M_{\tilde q}$.  In each case, we construct the effective Lagrangian
at the matching scale and compute the renormalization group equations
for the Wilson coefficients.  In Sec.~\ref{num} we present the results
of our numerical analysis.  We find that the QCD corrections tighten
the limits on the squark mass splittings by more than a factor of two.
(Of course, our results are subject to the usual uncertainties associated
with the hadronic matrix elements.)  We reserve Sec.~\ref{con} for our
conclusions.

\section{QCD corrections to the effective Lagrangian}
\label{form}

\subsection{\it General formalism and conventions}

To analyze a low energy physical process in a theory with several mass
scales, it is useful to construct an effective field theory, in which
particles with masses greater than the scale of interest are integrated
out \cite{review}.  This gives rise to an effective Lagrangian, which can
be written as follows,
\begin{equation}
{\cal L}_{\rm eff} = \sum_A C_A(\mu) {\cal O}_A(\mu) ,
\end{equation}
where $\mu$ is the renormalization scale, and
the operators ${\cal O}_A$ involve only low energy fields.  The
Wilson coefficients $C_A$ are obtained by matching S-matrix elements
in the full and effective theories at the threshold (matching) scale.
The heavy particles modify the ultraviolet behavior of the low energy
theory, but the infrared physics is the same in both.

In any realistic calculation, the Wilson coefficients and operators
must be evolved from the matching scale to the scale associated with the
low energy process under consideration.  The renormalization group
equations are given by\footnote{We do our calculation in the
${\overline {\rm MS}}$-scheme.}
\begin{equation}
\mu{d{\cal O}_A\over d\mu} = -\gamma_{AB} {\cal O}_B,\qquad
\mu{dC_A\over d\mu} = (\gamma^T)_{AB} C_B  ,
\end{equation}
where $\gamma$ is the anomalous dimension matrix.  Using the one-loop
RGE for the strong coupling,
\begin{equation}
\mu{dg_s\over d\mu} = \beta_1 g_s^3 ,
\end{equation}
where
\begin{equation}
\beta_1 = {1\over 16\pi^2}
\biggl(-11+{2\over 3}n_q+{1\over 6}n_{\tilde q}+2n_{\tilde g}\biggr)
\label{beta}
\end{equation}
and $n_q, n_{\tilde q}, n_{\tilde g}$ are the number of active quark,
squark and gluino flavors, respectively, one can find a formal solution
for the Wilson coefficients at any scale $\mu$,
\begin{equation}
C(\mu) = \exp\Biggl[{\gamma^T\over 2\beta_1 g_s^2}
\log\biggl({\alpha_s(\mu)\over\alpha_s(M)}\biggr)\Biggr]\ C(M) .
\label{solu}
\end{equation}
Here $M$ is the matching scale and $\gamma$ is the one-loop anomalous
dimension matrix, which is of order $g_s^2$.  This procedure resums the
leading logarithmic QCD corrections.

In the following subsections, we will use the effective field theory
formalism to construct the leading-logarithmic QCD-corrected $\Delta S
= 2$ effective Lagrangian in the low energy limit of the MSSM.  We will
consider models with the following hierarchies between the gluino and
first two generation squark masses:
$M_{\tilde g} \simeq M_{\tilde q}$,
$M_{\tilde g} \ll M_{\tilde q}$ and
$M_{\tilde g} \gg M_{\tilde q}$.

\subsection{$M_{\tilde g} \simeq M_{\tilde q}$}

When $M_{\tilde g} \simeq M_{\tilde q}$, we choose to integrate out the
squarks and gluino at the scale $M_{\rm SUSY}$, which we define to be the
geometric mean of the squark and gluino masses, $M_{\rm SUSY} =
\sqrt{M_{\tilde g}M_{\tilde q}}$.  The supersymmetric contribution to
the $\Delta S = 2$ effective Lagrangian is then
\begin{equation}
{\cal L}_{\rm eff} =
{\alpha_s^2(M_{\rm SUSY})\over 216 M^2_{\tilde q}}\ \sum_A
C_A(\mu) {\cal O}_A(\mu),
\label{full}
\end{equation}
where we have defined the operators
\begin{eqnarray}
&& {\cal O}_1 = {\overline d}_L^i\gamma_\mu s_L^i
{\overline d}_L^j \gamma^\mu s_L^j ,\nonumber\\[2mm]
&& {\cal O}_2 = {\overline d}_R^i s_L^i
{\overline d}_R^j s_L^j ,\nonumber\\[2mm]
&& {\cal O}_3 = {\overline d}_R^i s_L^j
{\overline d}_R^j s_L^i ,\\[2mm]
&& {\cal O}_4 = {\overline d}_R^i s_L^i
{\overline d}_L^j s_R^j ,\nonumber\\[2mm]
&& {\cal O}_5 = {\overline d}_R^i s_L^j
{\overline d}_L^j s_R^i ,\nonumber
\end{eqnarray}
plus other operators ${\tilde {\cal O}}_{1,2,3}$, with the obvious exchanges
$L \leftrightarrow R$ in ${\cal O}_{1,2,3}$.  In these expressions, the
superscripts $i, j$ are $SU(3)$ color indices.  All other operators with
the correct Lorentz and color structure can be related to these
by operator identities and Fiertz rearrangements.

At the matching scale $M_{\rm SUSY}$, we determine the Wilson
coefficients to be
\begin{eqnarray}
&&
C_1(M_{\rm SUSY}) = (24 x f_6(x) +
66 {\tilde f}_6(x))\,\delta_{LL}^2 ,\nonumber\\[2mm]
&&
C_2(M_{\rm SUSY}) = 204 x f_6(x) \,
\delta^2_{RL} ,\nonumber\\[2mm]
&&
C_3(M_{\rm SUSY}) = - 36 x f_6(x)\, \delta^2_{RL} ,
\label{wilsonc}\\[2mm]
&&
C_4(M_{\rm SUSY}) = (504 x f_6(x) -72
{\tilde f}_6(x))\,\delta_{LL}\delta_{RR}
-132 {\tilde f}_6(x)\,\delta_{LR}\delta_{RL} ,\nonumber\\[2mm]
&&
C_5(M_{\rm SUSY}) = (24 x f_6(x) + 120
{\tilde f}_6(x))\,\delta_{LL}\delta_{RR}
- 180 {\tilde f}_6(x)\,\delta_{LR}\delta_{RL} ,\nonumber
\end{eqnarray}
where $x = M^2_{\tilde g}/M^2_{\tilde q}$ and the $\delta$'s come
from insertions of the squark mass matrix.  The functions $f_6$
and ${\tilde f}_6$ \cite{hagelin}
\begin{eqnarray}
&& f_6(x) = {6 (1+3x)\log x + x^3 - 9 x^2 - 9x +17\over 6 (x-1)^5} ,
\nonumber\\[2mm]
&& {\tilde f}_6(x) = {6 x (1+x) \log x - x^3 - 9 x^2 + 9x +
1\over 3 (x-1)^5}
\end{eqnarray}
arise from momentum integrals in the gluino-squark box diagrams.
They have the limits
\begin{eqnarray}
x f_6(x) \rightarrow {\cal O}(x), &&
   \tilde f_6(x)\rightarrow -{1\over3} \qquad \quad (x\ll 1) \nonumber\\[2mm]
x f_6(x) \rightarrow {1\over 6x}, &&
   \tilde f_6(x)\rightarrow {\cal O}(x^{-2}) \qquad  (x\gg 1) ,
\label{limits}
\end{eqnarray}
so the leading behavior of the effective Lagrangian goes as $1/M_{\tilde
q}^2$ for  $x\ll 1$ and $1/M_{\tilde g}^2$ for $x\gg 1$.  The values of
the  coefficients $\tilde C_{1,2,3}$ are obtained by replacing $L
\leftrightarrow R$ in eq.~(\ref{wilsonc}).  Note that our result disagrees
with that of Ref.~\cite{hagelin}, but confirms that of Ref.~\cite{ggms}.

The operator ${\cal O}_1$ is of $V-A$ type.  Its anomalous dimension
is well known \cite{oldpaper}:
\begin{equation}
\gamma ({\cal O}_1) = {g_s^2\over 4\pi^2} .
\end{equation}
The one-loop anomalous dimensions for the other operators are
\begin{eqnarray}
&&\gamma ({\cal O}_2{\cal O}_3) = {g_s^2\over 12\pi^2}\left(\begin{array}{cc}
-7 & 1\\[2mm]
4 & 8
\end{array}\right)  ,  \\[2mm]
&&\gamma ({\cal O}_4{\cal O}_5) = {g_s^2\over 8\pi^2} \left(\begin{array}{cc}
-8 & 0\\[2mm]
-3 & 1
\end{array}
\right) .
\end{eqnarray}
With these anomalous dimensions, we evolve the Wilson
coefficients to the scale $\mu_{\rm had}$, where the hadronic
observables are defined.  We find
\begin{eqnarray}
&&C_1(\mu_{\rm had}) = \eta_{1} C_1 (M_{\rm SUSY}),\nonumber\\[2mm]
&&C_2(\mu_{\rm had}) = \eta_{22} C_2 (M_{\rm SUSY})
+\eta_{23} C_3 (M_{\rm SUSY}) ,\nonumber \\[2mm]
&&C_3(\mu_{\rm had}) = \eta_{32}
C_2 (M_{\rm SUSY}) +\eta_{33} C_3 (M_{\rm SUSY}) ,\nonumber\\[2mm]
&&C_4(\mu_{\rm had}) = \eta_4 C_4(M_{\rm SUSY})
+{1\over3} (\eta_4-\eta_5) C_5(M_{\rm SUSY}) ,\nonumber\\[2mm]
&&C_5(\mu_{\rm had}) = \eta_{5} C_5 (M_{\rm SUSY}),
\end{eqnarray}
where
\begin{eqnarray}
&&\eta_1 =
\biggl({\alpha_s(m_c)\over \alpha_s(\mu_{\rm had})}\biggr)^{6/
27}
\biggl({\alpha_s(m_b)\over \alpha_s(m_c)}\biggr)^{6/25}
\biggl({\alpha_s(m_t)\over \alpha_s(m_b)}\biggr)^{6/23}
\biggl({\alpha_s(M_{\rm SUSY})\over \alpha_s(m_t)}\biggr)^{6/21},
\nonumber\\[2mm]
&&\eta_{22} =  0.983 \eta_2 + 0.017 \eta_3,\ \  \eta_{23} =
-0.258 \eta_2 + 0.258 \eta_3,
\nonumber\\[2mm]
&&\eta_{32} =
-0.064 \eta_2 + 0.064 \eta_3,\ \  \eta_{33} =
0.017 \eta_2 + 0.983 \eta_3,\nonumber\\[2mm]
&&\eta_2 = \eta_1^{-2.42}, \ \ \eta_3 = \eta_1^{2.75},\ \
\eta_4 = \eta_1^{-4}, \ \ \eta_5 = \eta_1^{{1/2}} .
\label{eta1}
\end{eqnarray}

\vspace{2mm}
\subsection{$M_{\tilde g} \ll M_{\tilde q} \quad (x \ll 1)$}

\begin{figure}[t]
\epsfysize=3.2in \epsffile[0 230 660 570]{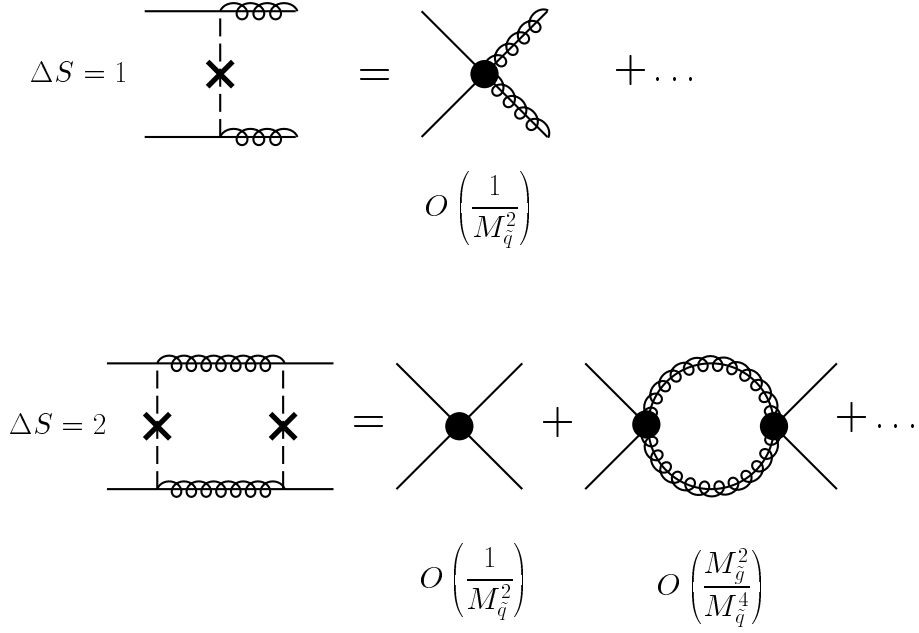}
\begin{center}
\parbox{5.5in}{
\caption[]{The matching of the S-matrices in the full and effective
theories, at the scale $M_{\tilde q}$,  for the case of $M_{\tilde g}
\ll M_{\tilde q}$.  At one loop, the $\Delta S=1$ operator generates
a subleading $\Delta S=2$ operator, of order $M_{\tilde g}^2/M_{\tilde
q}^4$.
\label{sq}}}
\end{center}
\end{figure}

When $M_{\tilde g} \ll M_{\tilde q}\ (x \ll 1)$, we proceed in
two steps.  We first integrate out the heavy squarks at their
own mass scale, $M_{\tilde q}$ (see Fig.~\ref{sq}).  This gives an
effective Lagrangian with $\Delta S=1$ and $\Delta S=2$ four-fermion
operators.  The $\Delta S=1$ operators are of the form
\begin{equation}
{g_s^2\over M^2_{\tilde q}} (T^aT^b)_{ij} {\overline d}^i_R {\tilde g}^a
 {\overline{\tilde g}}^b s^j_L ,
\end{equation}
together with similar operators with the obvious exchanges $L\leftrightarrow
R$ and/or $d\leftrightarrow s$.  The $\Delta S=1$ operators give rise
to one-loop $\Delta S=2$ diagrams which are suppressed by $M_{\tilde g}^2
/M_{\tilde q}^4$, so we ignore them in what follows.  The $\Delta S=2$
terms in the effective Lagrangian are given by
\begin{eqnarray}
{\cal L}_{\rm eff} &=& {\alpha_s^2(M_{\tilde q})\over 216 M_{\tilde q}^2}
 \biggl\{-22\delta^2_{LL}\,{\cal O}_1-22\delta^2_{RR}\,{\tilde {\cal O}}_1
+\delta_{LL}\delta_{RR}\,(24{\cal O}_4 - 40 {\cal O}_5) \nonumber\\[2mm]
&&+\delta_{LR}\delta_{RL}\,(44{\cal O}_4 + 60 {\cal O}_5)\biggr\} ,
\label{leff}
\end{eqnarray}
where the matching is performed at the scale $M_{\tilde q}$.

To construct the effective Lagrangian, these $\Delta S = 2$ operators must
first be evolved to the gluino scale, and then to the hadronic
scale.\footnote{Integrating out the gluinos does not induce any new $\Delta
S = 2$ operators.  However, it changes the beta function for the strong
coupling.
We have assumed that the third generation squarks have masses of order
$M_{\tilde g}$, in which case the one-loop beta function coefficient
is $-13/3$ between $M_{\tilde g}$ and $M_{\tilde q}$.}  The
one-loop anomalous dimensions for operators ${\cal O}_{1,4,5}$ and ${\tilde
{\cal O}}_1$ were calculated previously, so the leading $\Delta S=2$ effective
Lagrangian at the hadronic scale can be obtained by scaling eq.~(\ref{leff}).
We find
\begin{eqnarray}
{\cal L}_{\rm eff} &=& {\alpha_s^2(M_{\tilde q})\over 216
M_{\tilde q}^2}
 \biggl\{-22\delta^2_{LL}\kappa_1\,{\cal O}_1
-22\delta^2_{RR}\kappa_1\,{\tilde {\cal O}}_1\nonumber\\
&&+\delta_{LL}\delta_{RR}\,({8\over3}(4\kappa_4+5\kappa_5){\cal O}_4
- 40\kappa_5{\cal O}_5)\nonumber\\[2mm]
&&+\delta_{LR}\delta_{RL}\,((64\kappa_4-20\kappa_5){\cal O}_4
+ 60\kappa_5{\cal O}_5)\biggr\} ,
\end{eqnarray}
where
\begin{eqnarray}
&&\kappa_1 = \biggl({\alpha_s(m_c)\over \alpha_s(\mu_{\rm had})}\biggr)
^{6/ 27}
\biggl({\alpha_s(m_b)\over \alpha_s(m_c)}\biggr)^{6/25}
\biggl({\alpha_s(m_t)\over \alpha_s(m_b)}\biggr)^{6/23}
\biggl({\alpha_s(M_{\tilde g})\over \alpha_s(m_t)}\biggr)^{6/21}
\biggl({\alpha_s(M_{\tilde q})\over
\alpha_s(M_{\tilde g})}\biggr)^{6/13},\nonumber\\[2mm]
&&\kappa_4 = \kappa_1^{-4}, \qquad   \kappa_5 = \kappa_1^{{1/2}} .
\end{eqnarray}

\subsection{$M_{\tilde g} \gg M_{\tilde q} \quad (x\gg 1)$}

\begin{figure}[t]
\epsfysize=3.2in \epsffile[0 230 660 570]{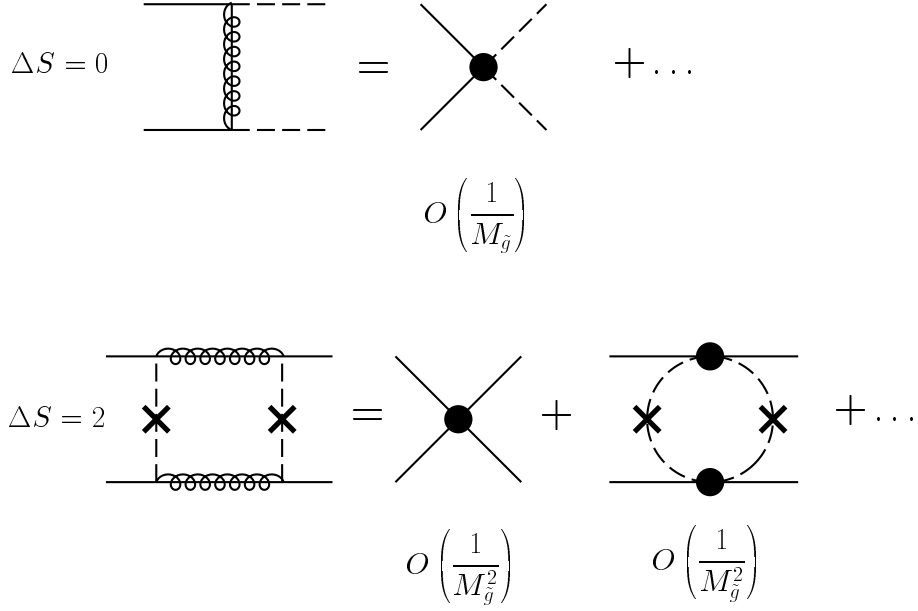}
\begin{center}
\parbox{5.5in}{
\caption[]{The matching of the S-matrices in the full and effective
theories, at the scale $\mu=M_{\tilde g}$, for the case $M_{\tilde g}
\gg M_{\tilde q}$.  The $\Delta S=0$ operator generates a leading
$\Delta S=2$ operator, whose coefficient exactly matches the $x \gg 1$
limit of the corresponding box diagram.  This implies that the Wilson
coefficient of the first $\Delta S=2$ operator on the right-hand side
is zero.
\label{gl}}}
\end{center}
\end{figure}

When $M_{\tilde g} \gg M_{\tilde q}\ (x\gg 1)$, we must again proceed
in two steps.  We first integrate out the gluino at $M_{\tilde g}$
(see Fig.~\ref{gl}).  The effective Lagrangian between $M_{\tilde q}$
and $M_{\tilde g}$,
\begin{equation}
{\cal L}_{\rm eff} = \sum_A D_A(\mu) {\cal Q}_A(\mu),
\end{equation}
contains the following leading order operators
\begin{eqnarray}
&&{\cal Q}_1 = {\overline d}_R^i s_L^i {\tilde d}^{j}_R
 {\tilde s}_L^{j*} ,\nonumber\\[2mm]
&&{\cal Q}_2 = {\overline d}_R^i s_L^j {\tilde d}^{i}_R
 {\tilde s}_L^{j*} ,\nonumber\\[2mm]
&&{\cal Q}_3 = {\overline s^c}_L^i s_L^j {\tilde s}^{j*}_L
 {\tilde s}_L^{i*},
\end{eqnarray}
as well as other operators with obvious exchanges, $L\leftrightarrow R$
and/or $s\leftrightarrow d$. The superscript $c$ in ${\cal Q}_3$ stands
for the charge conjugated field.

By matching at the scale $M_{\tilde g}$, we determine the Wilson
coefficients to be
\begin{equation}
D_1 = -3 D_2 = 6 D_3 = -{g_s^2(M_{\tilde g})\over M_{\tilde g}} .
\end{equation}
The one-loop anomalous dimensions are
\begin{equation}
\gamma ({\cal Q}_1{\cal Q}_2) = {g_s^2\over 8\pi^2}
\left(\begin{array}{cc}
-8 & 0\\[2mm]
-{3\over 2} & -{7\over 2} \end{array}\right) , \qquad
  \gamma ({\cal Q}_3) = -{3 g_s^2\over 8\pi^2}.
\end{equation}
Therefore the Wilson coefficients $D_A$ $(A=1,2,3)$ at the scale
$M_{\tilde q}$ are given by:
\begin{eqnarray}
D_1 (M_{\tilde q}) &=& \varepsilon_1 D_1(M_{\tilde
g})+{1\over3}(\varepsilon_1-\varepsilon_2) D_2(M_{\tilde g}) \nonumber\\
D_A (M_{\tilde q}) &=& \varepsilon_A D_A(M_{\tilde g}),\qquad A=2,3
\end{eqnarray}
where
\begin{equation}
\varepsilon_1 = \biggl({\alpha_s(M_{\tilde g})\over
\alpha_s(M_{\tilde q})}\biggr)^{-{8/ 5}}, \qquad
  \varepsilon_2 = \varepsilon_1^{{7/ 16}}, \qquad
  \ve_3 = \varepsilon_1^{{3/ 8}}.
\end{equation}

We now evolve the effective Lagrangian to the scale $M_{\tilde q}$,
where we integrate out all the squarks.  At that scale, the
$\Delta S=2$ effective Lagrangian is given by
\begin{eqnarray}
{\cal L}_{\rm eff}&=&
{\alpha_s^2(M_{\tilde g})\over 216 M_{\tilde g}^2}
\Biggl\{ \delta_{LL}\delta_{RR}\biggl({4\over3}(64\ve^2_1-\ve_2^2){\cal O}_4
+ 4 \ve_2^2 {\cal O}_5\biggr)\nonumber\\[2mm]
&&\qquad\qquad + \biggl[ \delta_{RL}^2 \biggl(
({2\over3}(64\ve^2_1-\ve_2^2)-8\ve_3^2){\cal O}_2
+ (2\ve_2^2 -8\ve_3^2) {\cal O}_3\biggr) \\[2mm]
&&\qquad\qquad\qquad + 4\ve_3^2 \delta^2_{LL} {\cal O}_1
+ (L\leftrightarrow R, {\cal O}\rightarrow\tilde{\cal O})  \biggr]
\Biggr\} .\nonumber
\end{eqnarray}
This effective Lagrangian matches the $M_{\tilde q} \ll M_{\tilde g}$
limit of the full theory.

Finally, this Lagrangian must be evolved to the hadronic scale.  We
find
\begin{eqnarray}
{\cal L}_{\rm eff}&=&{\alpha_s^2(M_{\tilde g})\over 216 M_{\tilde g}^2}
\Biggl\{
 \delta_{LL}\delta_{RR}\biggl(
{4\over3}(64\ve^2_1\eta'_4-\ve_2^2\eta'_5)\,{\cal O}_4
+ 4\ve_2^2\eta'_5 \,{\cal O}_5\biggr)\nonumber\\[2mm]
&&\qquad\qquad + \biggl[ \delta_{RL}^2 \biggl(
({2\over3}(64\ve^2_1-\ve_2^2)-8\ve_3^2)\eta'_{22}
+ (2\ve_2^2 -8\ve_3^2)\eta'_{23}\biggr){\cal O}_2\nonumber\\[2mm]
&&\qquad\qquad  + \delta_{RL}^2 \biggl(
({2\over3}(64\ve^2_1-\ve_2^2)-8\ve_3^2)\eta'_{32}
+ (2\ve_2^2-8\ve_3^2)\eta'_{33} \biggr){\cal O}_3\nonumber\\[2mm]
&&\qquad\qquad\qquad\qquad
+ 4\ve_3^2\eta'_1 \delta^2_{LL}\, {\cal O}_1
+(L \leftrightarrow R,{\cal O}\rightarrow\tilde{\cal O})  \biggr] \Biggr\} ,
\end{eqnarray}
where the $\eta'$'s have the same form as the $\eta$'s, with
the scale $M_{\rm SUSY}$ replaced by $M_{\tilde q}$ in
eq.~(\ref{eta1}).

\section{Numerical results}\label{num}

In this section, we will use the $\Delta S = 2$ effective Lagrangians to
compute the supersymmetric corrections to the $K$-meson mass difference,
\begin{equation}
\Delta m_K = 2 \Re \langle K|{\cal L}_{\rm eff}|{\overline K}\rangle .
\end{equation}
We shall see that the measured value, $\Delta m_K = 3.5 \times 10^{-12}$
MeV, places tight limits on the parameters $\delta$.

Note that the $\Delta S = 2$ effective Lagrangians can also be used to
compute the supersymmetric contribution to the CP-violating parameter
$\epsilon_K$,
\begin{equation}
|\epsilon_K| = {\Im \langle K|{\cal L}_{\rm eff}|{\overline K}\rangle
\over 2 \Re \langle K|{\cal L}_{\rm eff}|{\overline K}\rangle} .
\end{equation}
If the phases of the $\delta$'s are of order one, the experimental limits
on $\epsilon_K$ place very tight constraints on the squark mass
matrices (see, for example, Ref.~\cite{pokmis}).

\begin{table}[ht]
\begin{center}
\renewcommand{\arraystretch}{2}
\begin{tabular}{|c|c|c|c|c|} \hline
\multicolumn{1}{|c|}{$x$}
&
\multicolumn{2}{c|}{$\sqrt{|\Re\delta^2_{LL}|}$}
&
\multicolumn{2}{c|}{$\sqrt{|\Re\delta^2_{LR}|}$}
\\ \hline
$0.3$
& $2.3\times 10^{-2}$ & $3.0\times 10^{-2}$
& $5.5\times 10^{-3}$ & $2.8\times 10^{-3}$
\\ \hline
$1.0$
& $5.0\times 10^{-2}$ & $6.6\times 10^{-2}$
& $6.3\times 10^{-3}$ & $3.2\times 10^{-3}$\\ \hline
$4.0$
& $0.12$ & $0.16$
& $9.2\times 10^{-3}$ & $4.6\times 10^{-3}$\\ \hline
\multicolumn{1}{|c|}{ }
&
\multicolumn{2}{c|}{$\sqrt{|\Re\delta_{LL}\delta_{RR}|}$}
&
\multicolumn{2}{c|}{$\sqrt{|\Re\delta_{LR}\delta_{RL}|}$}\\ \hline
$0.3$
&$3.1\times 10^{-3}$ & $1.0\times 10^{-3}$
& $4.2\times 10^{-3}$ & $1.4\times 10^{-3}$ \\ \hline
$1.0$
& $3.6\times 10^{-3}$ & $1.2\times 10^{-3}$
& $7.2\times 10^{-3}$ & $2.4\times 10^{-3}$ \\ \hline
$4.0$
& $5.3\times 10^{-3}$ & $1.7\times 10^{-3}$
& $1.7\times 10^{-2}$ & $5.6\times 10^{-3}$\\ \hline
\end{tabular}
\parbox{5.5in}{
\caption[]{Limits on the $\Re\delta$'s, for squark mass $M_{\tilde q}=500$
GeV and different $x={M^2_{\tilde g}/M^2_{\tilde q}}$.  In each case,
the left (right) numbers are limits without (with) leading order
QCD corrections. }}
\end{center}
\end{table}

In each of these expressions, the effective Lagrangian ${\cal L}_{\rm eff}$
contains the operators ${\cal O}_1$ through ${\cal O}_5$, with coefficients
evaluated at the hadronic scale, which we define to be the scale where
$\alpha_s = 1$.   We evaluate the hadronic matrix elements and find
\begin{eqnarray}
&&\langle K|{\cal O}_1|{\ol K} \rangle = {1\over 3}m_K f_K^2 B_1 ,
\nonumber\\[2mm]
&&\langle K|{\cal O}_2|{\ol K} \rangle = -{5\over 24}\biggl(
{m_K\over m_s+m_d}\biggr)^2 m_K f_K^2 B_2 ,\nonumber\\[2mm]
&&\langle K|{\cal O}_3|{\ol K} \rangle = {1\over 24}\biggl(
{m_K\over m_s+m_d}\biggr)^2 m_K f_K^2 B_3 ,\\[2mm]
&&\langle K|{\cal O}_4|{\ol K} \rangle = \Biggl[{1\over 24}+{1\over 4}
\biggl(
{m_K\over m_s+m_d}\biggr)^2\Biggr]m_K f^2_K B_4 ,\nonumber\\[2mm]
&&\langle K|{\cal O}_5|{\ol K} \rangle = \Biggl[{1\over 8}+{1\over 12}
\biggl(
{m_K\over m_s+m_d}\biggr)^2\Biggr]m_K f^2_K B_5 .\nonumber
\end{eqnarray}
The coefficients $B_{1-5}$ characterize the long-distance hadronic
physics; we take $B_i = 1$, as determined in the vacuum insertion
approximation.

\begin{figure}[t]
\epsfysize=3.2in \epsffile[0 180 660 535]{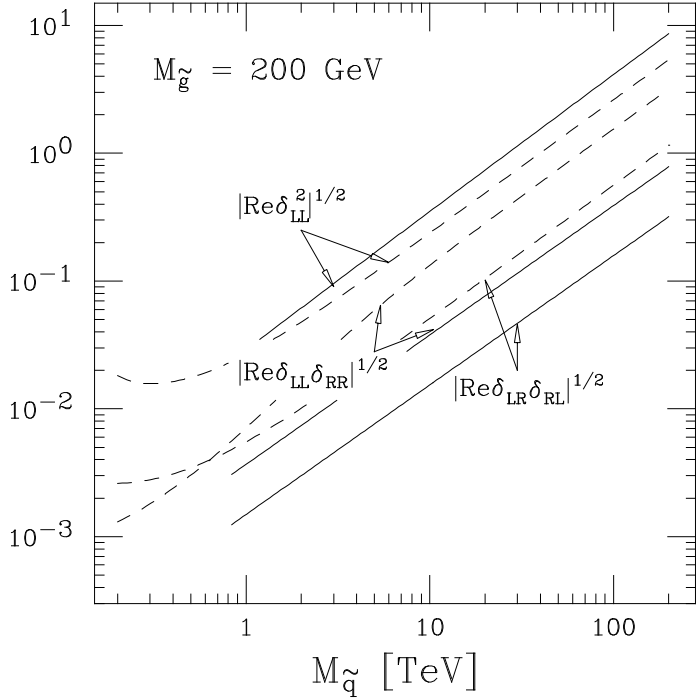}
\begin{center}
\parbox{5.5in}{
\caption[]{Limits on the $\Re \delta$'s as a function of the common
first two generation squark mass, $M_{\tilde q}$, for a light gluino
mass of $M_{\tilde g} = 200$ GeV.  The solid lines describe our
effective field theory result.  The dashed lines correspond to decoupling
the supersymmetric particles at $M_{\rm SUSY}= {\sqrt {M_{\tilde q}
M_{\tilde g}}}$, without including the leading order QCD corrections.
\label{fig1}}}
\end{center}
\end{figure}

In the rest of this section, we use these results to illustrate the size
of the leading order QCD corrections to $K$-${\overline K}$ mixing, for
each of the three cases discussed above.  For our numerical work, we
take $f_K = 160$ MeV, $m_K = 498$ MeV  \cite{pdg} and the current mass
$m_s = 150$ MeV.\footnote{This mass is not well determined; see
Ref.~\cite{strange} for recent analyses.  Decreasing $m_s$ to 100 MeV
tightens the bounds by at most 30\%.}
We use the strong coupling as determined from electroweak
measurements, $\alpha_s (M_Z) = 0.118$  \cite{pdg}.  This gives
$\alpha_s(m_c) = 0.35$ and $\alpha_s (m_b) = 0.22$.

We first consider the case $M_{\tilde g} \simeq M_{\tilde q}$.  The
QCD-corrected $\Delta S=2$ effective Lagrangian gives rise to the limits
shown in Table 1.  In the table, we have taken the first two generation
squark mass $M_{\tilde q}= 500$ GeV and varied $x = M^2_{\tilde g}/
M^2_{\tilde q}$.  The left (right) numbers correspond to the limits
without (with) the leading order QCD corrections.\footnote{Our numbers
agree with those of Ref.~\cite{ggms} if we follow their procedure
and decouple the squarks and gluino at $M_Z$ and neglect the QCD
corrections.}

The most stringent limit in the table comes from $\sqrt{|
\Re\delta_{LL}\delta_{RR}|}$.  For a given mass splitting, we see
the QCD corrections increase the squark lower bound by a factor of
three!  Note that the corrections are generally more than $70\%$, so
they cannot be ignored in computing limits on the squark masses.

\begin{figure}[t]
\epsfysize=3.2in \epsffile[0 180 660 535]{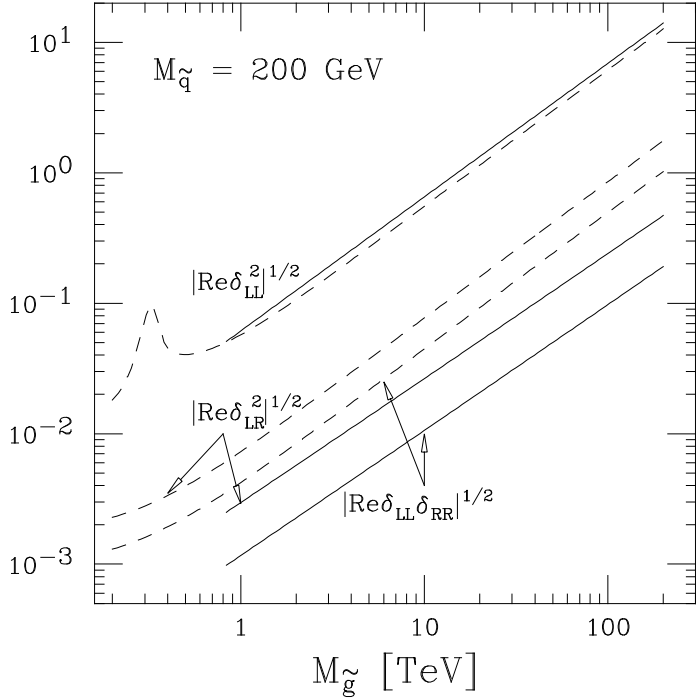}
\begin{center}
\parbox{5.5in}{
\caption[]{The same as Fig.~\ref{fig1}, as a function of $M_{\tilde g}$,
for light first two generation squarks, $M_{\tilde q}=200$ GeV.
\label{fig2}}}
\end{center}
\end{figure}

We next turn to models with very different squark and gluino masses.
In Fig.~\ref{fig1}, we consider the case of a light gluino, and
plot the limits on $\delta$'s versus the squark mass.  The solid
lines describe our results, while the dashed lines correspond to
decoupling the supersymmetric particles at $M_{\rm SUSY}={\sqrt
{M_{\tilde q} M_{\tilde g}}}$, with no QCD corrections.

{}From the figure we see that the QCD corrections tighten the most
stringent bounds by about $50\%$.  For $M_{\tilde g} = 200$ GeV,
with $\delta$ of order 1, the bound on the squark mass is $M_{\tilde q}
\gsim 200$ TeV.  Such a heavy squark can drive the third generation
squark/slepton mass squared negative if supersymmetry breaking is
transmitted by gravitational interactions \cite{mura}.

Alternatively, one can make the gluino heavy while keeping the
squarks light.  We show limits for this type of model
in Fig.~\ref{fig2}.  For $M_{\tilde q} = 200$ GeV, and $\delta$
of order 1, the bound on the gluino is $M_{\tilde g} \gsim 200$ TeV.
(Note, however, that in this scenario, the charginos and neutralinos
must also be heavy to suppress FCNC's from electroweak diagrams.)  In a
unified model, a large gluino mass drives up the squark masses
through their renormalization group evolution, so this scenario
can only be natural in models with TeV-scale supersymmetry breaking.

\section{Conclusion}\label{con}

In this paper we have calculated the leading order QCD corrections
to supersymmetric contributions to $K$-${\overline K}$ mixing.  We
find the corrections to be significant.  Indeed, for the case
$M_{\tilde q} \simeq M_{\tilde g}$, the QCD corrections increase
the lower bound on the first two generation squark masses by a factor
of three.

For the case $M_{\tilde q} \gg M_{\tilde g}$, we find that QCD
corrections also increase the squark lower bound.  This
exacerbates the naturalness problems associated with such a hierarchy,
and can destabilize the effective potential when supersymmetry
breaking is transmitted by supergravity interactions.
Similar results hold when $M_{\tilde q} \ll M_{\tilde g}$.  In each
case, the QCD corrections are important and must be included when
deriving constraints from FCNC processes.

We would like to thank A. Falk and M. Booth for very helpful
discussions, and K. Agashe and M. Graesser for finding a
mistake in a previous version of the paper.
This work was supported by the U.S. National Science
Foundation, grant NSF-PHY-9404057.

\bigskip

\bibliographystyle{unsrt}

\begin{thebibliography}{99}

\bibitem{longlist}
M.J. Duncan, \npb 221 (1983) 285;\ J.F. Donoghue, H.P. Nilles and D. Wyler,
\plb 128 (1983) 55;\ A. Bouquet, J. Kaplan and C.A. Savoy, \plb 148 (1984) 69;\
M.J. Duncan and J. Trampetic, \plb 134 (1984) 439;\ J.M.~Gerard, W.~Grimus,
A.~Raychaudhuri and G.~Zoupanos,
\plb 140 (1984) 349;\ J.M. Gerard, W. Grimus,
A. Masiero, D.V. Nanopoulos and A. Raychaudhuri,
\npb 253 (1985) 93;\ M. Dugan, B. Grinstein and L. Hall, \npb 255 (1985) 413;\
L.J. Hall, V.A. Kostelecky and S. Raby,
\npb 267 (1986) 415;\ Y. Nir, \npb 273 (1986) 567;\ G.C. Branco, G.C. Cho, Y.
Kizukuri and N. Oshimo,
\plb 337 (1994) 316.

\bibitem{list}
S. Bertolini, F. Borzumati, A. Masiero and G. Ridolfi,
\npb 353 (1991) 591;\ H. Anlauf, \npb 430 (1994) 245;\ P. Cho, M. Misiak and D.
Wyler, \prd 54 (1996) 3329;\ H. Baer and M. Brhlik, \prd 55 (1997) 3201.

\bibitem{old}
F.~Gabbiani and A.~Masiero, \npb 322 (1989) 235.

\bibitem{hagelin}
J.~Hagelin, S.~Kelley and T.~Tanaka, \mpla 8 (1993) 2737;
\npb 415 (1994) 293.

\bibitem{ggms}
E. Gabrielli, A. Masiero and L. Silvestrini,
\plb 374 (1996) 80; F. Gabbiani, E. Gabrielli, A. Masiero
and L. Silvestrini, \npb 477 (1996) 321.

\bibitem{GW}
F. Gilman and M. Wise, \prd 20 (1979) 2392;
\plb 93 (1980) 129; \prd 27 (1983) 1128;\ J. Flynn, \mpla 5 (1990) 877;\ A.
Datta, J. Fr\"ohlich and E.A. Paschos, \zpc 46 (1990) 63.

\bibitem{NLO}
A.J. Buras, M. Jamin and P.H. Weisz, \npb 347 (1990) 491;\ S. Herrlich and U.
Nierste, \npb 419 (1994) 292;
\prd 52 (1995) 6505; \npb 476 (1996) 27.

\bibitem{pokmis}
M. Misiak, S. Pokorski and J. Rosiek,
hep-ph/9703442, to appear in {\it `Heavy Flavors II'},
eds. A.J. Buras and M. Lindner,
World Scientific, Singapore.


\bibitem{review}
For reviews, see
H. Georgi, {\sl Weak Interactions and Modern Particle Theory},
Benjamin/Cummings, Menlo Park, CA (1984);
{\it Ann. Rev. Nucl. Part. Sci.} 43 (1994) 209.

\bibitem{oldpaper}
M. Gaillard and B.W. Lee, \prl 33 (1974) 108.

\bibitem{pdg}
Particle Data Group, \prd 54 (1996) 1.

\bibitem{strange}
T. Bhattacharya, R. Gupta and K.
Maltman, hep-ph/9703455; P. Colangelo, F. De Fazio, G. Nardulli and N.
Paver, hep-ph/9704249; SESAM Collaboration (N. Eicker {\it et al.}),
hep-lat/9704019.

\bibitem{mura}
N. Arkani-Hamed and H. Murayama, hep-ph/9703259.

\end{thebibliography}

\end{document}